\documentclass{iopart}
\usepackage{graphicx}
\usepackage{amssymb}
\begin{document}

\def\Dirac{i\partial\!\!\!\!/}
\def\DDirac{iD\!\!\!\!/}
\def\dirac{i\partial\!\!\!\!/-eA\!\!\!\!/}

\newcommand{\be}{\begin{eqnarray}}
\newcommand{\ee}{\end{eqnarray}}
\newcommand{\nn}{\nonumber}
\newcommand{\rparen}{({\bf r})}
\newcommand{\kparen}{({\bf k})}
\newcommand{\sumn}{\sum\limits}
\newcommand{\bfr}{{\bf r}}
\newcommand{\bfk}{{\bf k}}
\newcommand{\bfK}{{\bf K}}
\newcommand{\bfp}{{\bf p}}
\newcommand{\bfKpm}{{\bf K_{\pm}}}
\newcommand{\varpm}{{\varphi_{\pm}}}
\newcommand{\apm}{{a_{\pm}}}
\newcommand{\bpm}{{b_{\pm}}}
\newcommand{\ka}{{\kappa}}
\article[Dirac field in $AdS_3$]{}{Casimir energy for Dirac fields in $AdS_3$}

\date{\today}

\author{C.G. Beneventano}\address{Departamento de F\'{\i}sica -
Universidad Nacional de La Plata and IFLP- CONICET}
\author{E. M. Santangelo}\address{Departamento de F\'{\i}sica -
Universidad Nacional de La Plata and IFLP- CONICET}

\begin{abstract}

We study massless Dirac fields in a portion of $AdS_3$, where one of the boundaries coincides with the ``boundary at infinity'' of the Anti-de-Sitter space. We evaluate the vacuum energy arising when the local boundary conditions dictated by boundary chirality are imposed, in different combinations, at both one-branes.

\end{abstract}


\section{Introduction}\label{sect1}

Field theories in $2+1$ dimensions have been the subject of study from long ago (for a review see, for instance, \cite{dunne}), mainly due to their interesting topological properties, related to the existence of Chern-Simons gauge theories \cite{chern}. Although the relevance of these so-called planar theories in describing condensed matter phenomena was recognized from the beginning \cite{zhang,wilczek}, it became even more patent since the achievement of the experimental synthesis of graphene (a genuine two-dimensional electron system) \cite{science,nature1}.

Graphene is a bidimensional array of carbon atoms, packed in a honeycomb crystal structure (for a recent review see, for instance, \cite{neto}). From a theoretical point of view, the most remarkable feature of graphene is that, in a small momentum approximation, the charge carriers or {\sl quasi}--particles behave as two ``flavors" (to account for the
spin of the elementary constituents) of massless
relativistic Dirac particles in the two non--equivalent
representations of the Clifford algebra (corresponding to the two
non--equivalent vertices in the first Brillouin zone), with an effective ``speed of light"
about two orders of magnitude smaller than $c$ \cite{mele}.

The electronic properties of flat graphene sheets have been studied, in the field-theoretic approach, by many authors, including the authors of the present paper (see \cite{gus,BS1} and references therein).

However, both suspended and deposited-over-substrate samples, show corrugations (ripples)\cite{ripples}. This is one of the most intriguing properties of graphene, not entirely understood at present. Two main approaches are used to model such corrugations; they are based either on the presence of topological defects in the lattice \cite{sitenko} or on the theory of elasticity \cite{meyer}. In any case, a general relativity formalism can be applied by coupling the massless Dirac field to a curved background metric \cite{fernando}. In this paper, we will study a massless Dirac field in $2+1$ dimensional Anti de Sitter spacetime ($AdS_3$).

$AdS$ spaces have been intensively studied during the last decade, starting from the so-called $AdS-CFT$ conjecture \cite{malda} and the possibility of solving the hierarchy problem in brane-world scenarios \cite{lisa}. In particular, $AdS_3$ was studied, for instance, in \cite{teitel}.

More recently, an extra interest in the $AdS-CFT$ correspondence arose, since it was understood that it provides a tool to study condensed matter theories for superconductivity, superfluidity and the quantum Hall effect \cite{gubser} (see also \cite{adrian2} and references therein).

Boundary conditions are relevant when studying field theories in $AdS$ spaces (see, for example, \cite{bachelot}). For this reason, the Casimir effect \cite{casimir} for scalar fields in an $AdS_5$ background was treated by several authors \cite{emilio}. The main result of the present piece of work is the evaluation of the Casimir energy for a massless Dirac field in a portion of $AdS_3$, where one of the boundaries coincides with the ``boundary at infinity'' of this space.

The outline of the paper is as follows: In section \ref{sect2}, we give the differential expression of the Dirac operator in Poincar\'{e} coordinates, and find its general solutions. In section \ref{sect3}, the domain of the operator is restricted to functions satisfying particular local boundary conditions, and the corresponding energy modes are determined for the two nonequivalent combinations of such conditions at the boundaries. Section \ref{sect4} presents the calculation of the Casimir energy in both cases. Finally, section \ref{sect5} contains a discussion of the results.

\section{The Dirac operator in $AdS_3$ and its solutions}\label{sect2}

Fields of various spins, coupled to an $AdS_5$ background metric were studied, for instance, in \cite{grossman}. For Euclidean $AdS$, Dirac fields were treated in \cite{sfetsos}. Here, we study solutions of the Dirac equation in $2+1$ dimensions with the $AdS$ metric which, in Poincar\'{e} coordinates (for a definition see, for instance \cite{adrian}), is given by
\be
ds^2 = \left(\frac{\xi}{R}\right)^{-2}(-dt^2 +dx^2 + d\xi^2)\,.
\label{metric}\ee
Here, $R$ is the $AdS$ curvature radius. The warped coordinate $\xi$ is such that $0<\xi<\infty$, and the boundary of $AdS_3$ is given by $\xi =0$ plus one point at infinity \cite{boschi1}.

After choosing the convention $x:(x^{0}=t,x^{1}=x,x^{2}=\xi )$, the gamma matrices can be written in terms of the dreibeins and the flat gammas as
\be
\gamma ^{\mu}(x)=e^{\mu}_{j}(x) \gamma ^j \, .
\ee
By choosing $\gamma^0 = i \sigma _2$, $\gamma^1 = \sigma _1$ and $\gamma^2 = \sigma _3$, we get
\be
\gamma^0 (y)=-i \left(\frac{\xi}{R}\right)\sigma _2 \quad  \gamma^1 (y)= \left(\frac{\xi}{R}\right)\sigma _1 \quad \gamma^2 (y)= \left(\frac{\xi}{R}\right)\sigma _3 \, .\ee

Note that this amounts to choosing one of the two nonequivalent representations of the Clifford algebra existing in $2+1$ dimensions, as in any odd space-time dimension. We will comment on the effect of choosing the other nonequivalent representation in section \ref{sect5}.

From these space-time dependent Dirac matrices, together with the metric in equation (\ref{metric}), the components of the spinorial affine connection are determined as explained, for instance, in \cite{sucu}. They are given by
\be
\Gamma _0 (x)= - \frac {\xi^{-1}}{2} \sigma _1 \quad
\Gamma _1 (x)= - i\frac {\xi^{-1}}{2}\sigma _2 \quad
\Gamma _2 (x)=0 \, .
\ee
So, the Dirac equation is given by
\be
\left(\frac{\xi}{R}\right)\left[ \sigma _2 {\partial} _t + i \sigma _1 {\partial} _x -i \frac{\sigma _3}{\xi} \right]\Psi(t,x,\xi)=0 \,
\label{dirac1} .
\ee

After defining \be \Psi(t,x,\chi)=\xi \widetilde{\Psi}(t,x,\xi)\, , \label {tilde} \ee we have
\be
\left(\frac{\xi}{R}\right)^2\left[ \sigma _2 {\partial} _t + i \sigma _1 {\partial} _x + i\sigma _3  {\partial} _{\xi} \right]\widetilde{\Psi}(t,x,\xi )=0 \, .
\label{dirac2}\ee

Moreover, while the solutions in equation (\ref{dirac1}) must belong to $L^2\left(d^3x,\left(\frac{\xi}{R}\right)^{-2}\right)$, in order to have a well defined action (see, for instance, \cite{grossman}, where a discussion of the relevant scalar products, in the case of $AdS_5$ is presented), for $\widetilde{\Psi}$ such factor cancels, as can be seen from equation (\ref {tilde}). So, this last set of solutions must be square integrable with the flat metric.

In order to solve equation (\ref{dirac2}), we propose
\be
\widetilde{\Psi}(t,x,\xi )=e^{-i\,E\,t+i\,k_{x}\,x}\psi(\xi)=e^{-i\,E\,t+i\,k_{x}\,x}\left(\begin{array}{c}
                                                                                             \varphi(\xi)\\
                                                                                             \chi(\xi)
                                                                                           \end{array}
\right)
\,.\ee

\bigskip

We will distinguish two types of solutions:

\bigskip

i) Solutions corresponding to $E=\pm k_x$

For $E=k_x$, we have
\be
\psi=\left(\begin{array}{c}
             -2i\,c_1E \,\xi +c_2\\
             c_1
           \end{array}
\right)\label{sol1}\,,\ee
where $c_1$ and $c_2$ are arbitrary constants.

In turn, for $E=-k_x$, we have
\be
\psi=\left(\begin{array}{c}
             c_3\\
            -2i\,c_3E \,\xi +c_4
           \end{array}
\right)\label{sol2}\,,\ee
with $c_3$ and $c_4$ arbitrary constants.

\bigskip

ii) Solutions with $E^2 \neq {k_x}^2$, which are given by
\be
\psi=\left(\begin{array}{c}
            A \sin{(\lambda\,\xi)}+B \cos{(\lambda\,\xi)}\\
            \frac{i\,\lambda}{E+k_x}\left[A \cos{(\lambda\,\xi)}-B \sin{(\lambda\,\xi)}\right]
           \end{array}
\right)\label{sol3}\,,\ee
where $A$ and $B$ are arbitrary constants, and $\lambda=+\sqrt{E^2 - {k_x}^2}$ are the eigenvalues of the $\xi$- independent part of the differential operator. Note that, since we will impose boundary conditions at given values of the variable $\xi$, these are the eigenvalues of the boundary Dirac operator.

\section{Boundary conditions and energy modes}\label{sect3}

In order to complete the definition of the Dirac operator, we must determine its domain. We will consider a finite portion of $AdS_3$, with boundaries at $\xi=0$ and $\xi=R$. Note that $\xi=0$ is nothing but the so called ``boundary at infinity'' of the $AdS$ space. We will impose generalized bag boundary conditions at both boundaries. These local boundary conditions are known to define a well posed (elliptic \cite{gilkey}) boundary problem for the Dirac operator in Minkowski space-time \cite{obstruction}, and to represent a vanishing flow of current through the boundary, as first noted in the framework of the effective bag model for quark confinement \cite{mit}. In this particular setup, where both boundaries lie at definite values of the coordinate $\xi$, bag boundary conditions turn out to be the local ones dictated by boundary chirality \cite{wojce}.

Of the four possible combinations of such conditions, only two give rise to different Casimir energies. We will analyze each of the possible combinations in what follows.
\bigskip

\[\fl 1)\quad\left.\frac{1+{\sigma}_3}{2}\psi\right\rfloor_{\xi=0}=\left.\frac{1+{\sigma}_3}{2}
\psi\right\rfloor_{\xi=R}=0\,.\]

After imposing these boundary conditions, the only nontrivial solutions of type i) (see previous section) are those corresponding to $E_0=-k_x$. They adopt the form

\[
\psi=\left(\begin{array}{c}
             0\\
            c_4
           \end{array}
\right)\,.\]

As for the solutions of type ii), they are
\[\psi_n=A\left(\begin{array}{c}
             \sin{(\frac{n\,\pi}{R}\,\xi)}\\
            \frac{i\,n\,\pi}{R(E_n+k_x)}\,\cos{(\frac{n\,\pi}{R}\,\xi)}
           \end{array}
\right)\,,\]
with $E_n =\pm \sqrt{{(\frac{n\,\pi}{R})}^2 +k_x^2},\qquad n=1,2,...$.

\bigskip

\[\fl 2)\left.\frac{1-{\sigma}_3}{2}\psi\right\rfloor_{\xi=0}=\left.\frac{1-{\sigma}_3}{2}
\psi\right\rfloor_{\xi=R}=0\,,\]
which correspond to asking that the down component, $\chi$, of $\psi$ vanishes at both boundaries. It is very easy to check that for the solutions of type ii) the same energy modes are obtained as in $1)$, while the only non-vanishing solutions of type i) correspond to $E=k_x$. However, the contribution of these modes to the vacuum energy is the same as the one due to the modes with $E=-k_x$ in the previous case.

\bigskip

\[\fl 3)\quad\left.\frac{1+{\sigma}_3}{2}\psi\right\rfloor_{\xi=0}=\left.\frac{1-{\sigma}_3}{2}
\psi\right\rfloor_{\xi=R}=0\,.\]

In this case, all the solutions of type i) are trivial, and those corresponding to ii) are
\[ \psi_n=A\left(\begin{array}{c}
             \sin{\left(\frac{(n+ \frac{1}{2})\pi}{R}\,\xi\right)}\\
            \frac{i(n+ \frac{1}{2})\pi}{R(E_n+k_x)}\cos{\left(\frac{(n+ \frac{1}{2})\pi}{R}\,\xi\right)}
           \end{array}
\right)\,,\]
with $E_n =\pm \sqrt{{\left(\frac{(n+\frac{1}{2})\pi}{R}\right)}^2 +k_x^2},\qquad n=0,1,...$.

\[\fl 4)\quad\left.\frac{1-{\sigma}_3}{2}\psi\right\rfloor_{\xi=0}=\left.\frac{1+{\sigma}_3}{2}
\psi\right\rfloor_{\xi=R}=0\,.\]

In this case, exactly the same energy modes as in $3)$ arise.

\section{Casimir energies}\label{sect4}

In what follows, we will evaluate the vacuum energies \cite{plunien} corresponding to the boundary conditions 1) and 3) in the previous section, as defined in the framework of the zeta regularization \cite{dowker},
\be
\left. E_C =-\frac12 \left[\sum_{E>0} E^{-s}+\sum_{E<0} |E|^{-s}\right]\right\rfloor_{s=-1}\,.
\ee

Moreover, we will compactify the $x$-variable in the interval $0\leq x\leq \beta$, imposing antiperiodic boundary conditions on the Dirac fields. This will give as a result the allowed discrete values of $k_x=(l+\frac12 )\frac{2\pi}{\beta}$, with $l=-\infty,...,\infty $.

\bigskip

Boundary conditions 1): There are two contributions, $E_{i}$ and $E_{ii}$, to the vacuum energy. Their zeta-regularized expressions are

\be
\left.E_{i}=-\sum_{l=0}^{\infty}\left((l+\frac12 )\frac{2\pi}{\beta}\right)^{-s}\right\rfloor_{s=-1}=-\frac{2\pi}{\beta}\zeta_H (s=-1,\frac12)\,,
\label{e1}\ee
where $\zeta_H $ is the Hurwitz zeta function, and
\be
\left.E_{ii}=-\sum_{l=-\infty}^{\infty}\sum_{n=1}^{\infty}\left[(\frac{n\pi}{R})^2+\left[(l+\frac12 )\frac{2\pi}{\beta}\right]^2\right]^{-\frac{s}{2}}\right\rfloor_{s=-1}\,.
\label{e2}\ee

Note that the last expression converges for $\Re s>2$. By using the Mellin transform, the last expression can be written as
\be
\left.E_{ii}=-\frac{1}{\Gamma\left(\frac{s}{2}\right)}\int_0 ^{\infty}dt\,t^{\frac{s}{2}-1}\sum_{l=-\infty}^{\infty}\sum_{n=1}^{\infty}e^{-t\left[(\frac{n\pi}{R})^2+\left[(l+\frac12 )\frac{2\pi}{\beta}\right]^2\right]}\right\rfloor_{s=-1}.
\ee
Its analytical extension can be performed by the well known method based on the properties of  the Jacobi theta function \cite{nosotras} or, equivalently, by making use of \cite{klaus}
\be
\sum_{l=-\infty}^{\infty}e^{-t(l+c)^2}=\left(\frac{\pi}{t}\right)^{\frac12}\sum_{l=-\infty}^{\infty}
e^{-\frac{{\pi}^2}{t}l^2-2\pi i\, l c}\,.
\ee
After doing so, we get
\be
\nn E_{ii}&=&-\frac{\beta}{\sqrt{4\pi}}\frac{1}{\Gamma\left(\frac{s}{2}\right)}
\left\{2\sum_{l=1}^{\infty}\sum_{n=1}^{\infty}(-1)^l\int_0 ^{\infty}dt\,t^{\frac{s-1}{2}-1}e^{-t(\frac{n\pi}{R})^2-\frac{l^2 \beta^2}{4t}} \right.\\ &+& \left.\left.\sum_{n=1}^{\infty}\int_0 ^{\infty}dt\,t^{\frac{s-1}{2}-1}e^{-t(\frac{n\pi}{R})^2}\right\}\right\rfloor_{s=-1}.
\ee

Now, after performing the integral in the first term, and writing the second one as a Riemann zeta function ($\zeta_R$), we obtain
\be
\nn E_{ii}&=&-\frac{\beta}{\sqrt{4\pi}}\frac{1}{\Gamma\left(\frac{s}{2}\right)}
\left\{4\sum_{l=1}^{\infty}\sum_{n=1}^{\infty}(-1)^l \left(\frac{l\beta R}{2n\pi }\right)^{\frac{s-1}{2}}K_{\frac{s-1}{2}}\left(\frac{l\,n\pi \beta}{ R}\right) \right.\\ &+& \left.\left.\Gamma\left(\frac{s-1}{2}\right)(\frac{\pi}{R})^{1-s}\zeta_R (s-1)\right\}\right\rfloor_{s=-1}.
\ee

Finally, after using the reflection formula for the Riemann zeta function \cite{gradshteyn}, and taking into account the contribution from equation (\ref{e1}), the total Casimir energy is given by
\be \fl
E_C =-\frac{\pi}{12\beta}+\frac{\beta}{4\pi}\left\{8\pi\sum_{l,n=1}^{\infty}(-1)^l\frac{n }{l\beta R}K_1 \left(\frac{l\,n\pi \beta}{R}\right)+\frac{1}{2{R}^2}\zeta_R (3)\right\}\,.
\label{ec1}\ee

\bigskip

Boundary conditions 3): In this case, there is only one contribution to the Casimir energy since, as remarked in the previous section, no solution of type i) is left. So, we have
\be
\left.E_C =-\sum_{l=-\infty}^{\infty}\sum_{n=0}^{\infty}\left[\left[(n+\frac12)\frac{\pi}{R}\right]^2+\left[(l+\frac12 )\frac{2\pi}{\beta}\right]^2\right]^{-\frac{s}{2}}\right\rfloor_{s=-1}\,.\ee

The analytic extension of this expression can be performed following the same steps as in the previous calculation, to obtain
\be
\nn \fl E_{C}&=&-\frac{\beta}{\sqrt{4\pi}}\frac{1}{\Gamma\left(\frac{s}{2}\right)}
\left\{4\sum_{l=1}^{\infty}\sum_{n=0}^{\infty}(-1)^l \left(\frac{l\beta R}{2(n+\frac12 )\pi }\right)^{\frac{s-1}{2}}K_{\frac{s-1}{2}}\left(\frac{l(n+\frac12)\pi \beta}{R}\right) \right.\\ \fl &+& \left.\left.\Gamma\left(\frac{s-1}{2}\right)(\frac{\pi}{R})^{1-s}\zeta_H (s-1,\frac12 )\right\}\right\rfloor_{s=-1}.
\ee

The evaluation of the second term inside the curly brackets can be performed after using the relationship between the Hurwitz and Riemann zeta functions \cite{gradshteyn} and, then, applying the reflection formula for the last one. Thus, one gets
\be \fl
E_C =\frac{\beta}{4\pi}\left\{8\pi\sum_{l=1,n=0}^{\infty}(-1)^l\frac{(n+\frac12) }{l\beta R}K_1 \left(\frac{l\,(n+\frac12 )\pi \beta}{R}\right)-\frac{3}{8{R}^2}\zeta_R (3)\right\}\,.
\label{ec2}\ee

\section{Final comments and remarks}\label{sect5}

Equations (\ref{ec1}) and (\ref{ec2}) are the main result in this paper, i.e., the vacuum energy of a massless Dirac field in a portion  of $AdS_3$, with different combinations of chiral boundary conditions.

It is interesting to note that, in the limit $\beta \rightarrow \infty$, the vacuum energy per unit length reduces, for the boundary conditions of type $1)$ to
\be
\lim_{\beta \rightarrow \infty}\frac{E_C}{\beta}=\frac{1}{8\pi {R}^2}\zeta_R (3)\,.
\ee
which is positive.

Instead, for boundary conditions of type $3)$, the same limit gives as a result the negative contribution
\be
\lim_{\beta \rightarrow \infty}\frac{E_C}{\beta}=-\frac{3}{32\pi {R}^2}\zeta_R (3)\,.
\ee

As explained before, throughout or calculation we have considered one of the two nonequivalent representations of the gamma matrices. The other representation differs by a change in the sign of one of the gamma matrices. Take, for instance, ${\sigma}_1\rightarrow  -{\sigma}_1$ or, equivalently $k_x \rightarrow -k_x$. If the same boundary conditions are imposed, the eigen-energies $1)$ transform into the eigen-energies $2)$, while those in $3)$ remain unchanged. But, as already commented in the previous section, this has no consequence on the final result for the corresponding vacuum energies.

An open problem, of particular interest in the case of graphene, is the study of the finite-temperature properties of the present model. This would imply the calculation of the partition function in Euclidean AdS. Note that the boundary conditions used in this paper have been shown to define a well-posed boundary problem for the Dirac operator not only in Minkowski, but also in Euclidean space \cite{bgks}.

Finally, we remark that the (local) boundary conditions imposed in this work are far from being the only possible ones. For instance, nonlocal boundary conditions of the Atiyah-Patodi-Singer (APS) type \cite{aps} also define an elliptic boundary problem \cite{gilkey}. In fact, their use in the context of string theory was proposed in \cite{dima}.

\ack{We thank Adri\'{a}n R. Lugo for useful discussions and suggestions. This work was partially supported by Universidad Nacional de La Plata (Proyecto 11/X492), ANPCyT (PICT 909) and CONICET (PIP 1787).}

\end{document}